\documentclass[11pt]{article}  
\usepackage{menuproc}
\usepackage{cite}
\usepackage{epsfig}

\newcommand{\GeVc}{$({\rm GeV}/c)^2$}

\begin{document}
\titlematter[7 September 2001]{Analysis of pion electroproduction data}%
{R. A. Arndt, I. I. Strakovsky, and R. L. Workman}%
{Center for Nuclear Studies, Department of Physics,\\
     The George Washington University, Washington, DC 20052, U.S.A.}%
{A fit to the existing pion electroproduction data is presented.
This work builds upon our previous analyses of pion photoproduction
and elastic pion-nucleon scattering over the Delta resonance region.
We comment on the extraction of $E_{1+}^{3/2} /  M_{1+}^{3/2}$
(E2/M1) and $S_{1+}^{3/2} / M_{1+}^{3/2}$ (S2/M1) ratios, and note
that the E2/M1 ratio approaches, and possibly crosses, zero below a
$Q^2$ of 5~\GeVc.}
%

\begin{figure}[b!]
\centerline{%
\epsfig{file=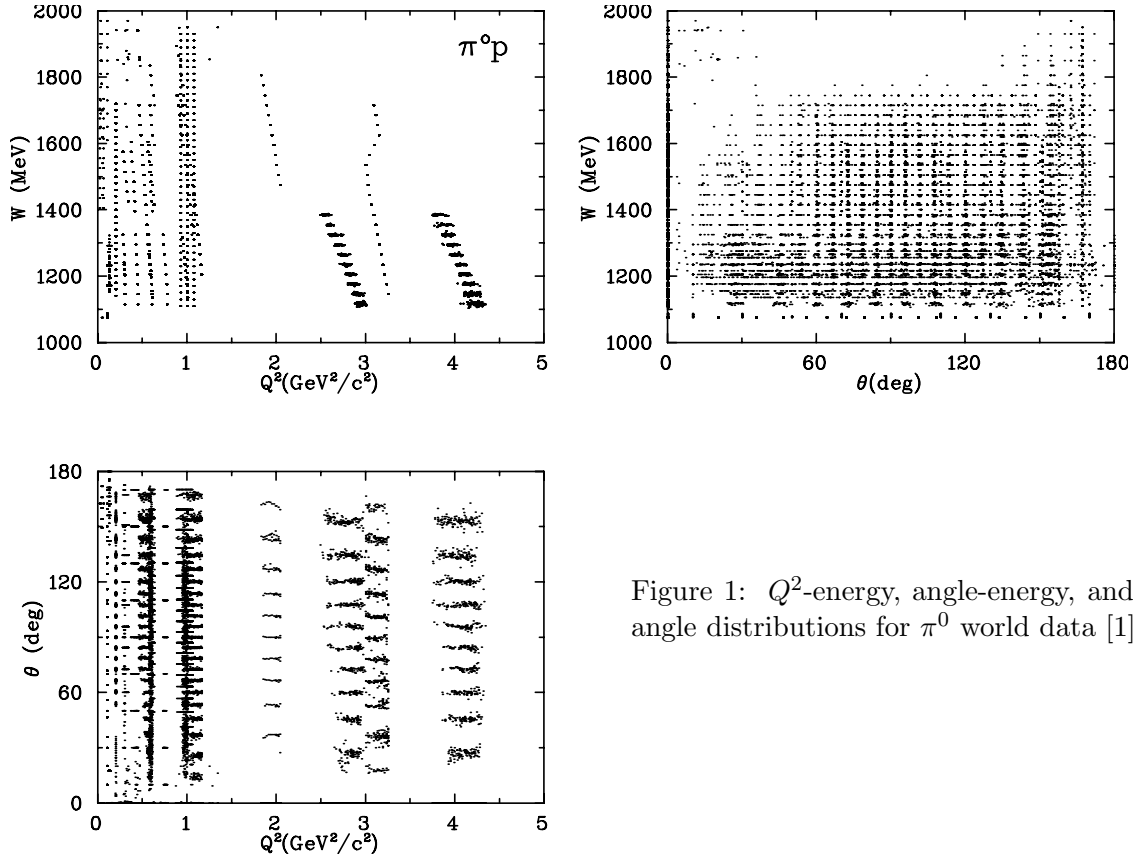,height=.9\textwidth,clip=,silent=,angle=90}}
\vspace{-8mm}
\hspace*{.55\textwidth}\raisebox{30mm}[0pt][0pt]{\parbox{.45\textwidth}{\caption[fig1]{\label{fig1}
$Q^2$-energy, angle-energy, and $Q^2$-angle distributions
for $\pi^0$ world data \protect\cite{SAID}.}}}
\end{figure}

\section{Introduction}

Over the last several years, we have assembled a database containing
the existing pion electroproduction data~\cite{SAID} and have made a
number of trial fits to this set, exploring possible extensions to
the methods we have applied to pion photoproduction~\cite{GW_photo}.
These efforts have intensified now that a flood of new and precise
data is becoming available from measurements performed at Jefferson
Lab, Mainz and Bonn.  Preliminary CLAS data (unpolarized
and beam polarization $\pi ^0$ and $\pi ^+$~\cite{CLASB} and double
polarization $\pi ^+$~\cite{CLASD}) will soon increase the database
size from approximately 10K to 30K points.  These new data were
taken at CM energies covering mainly the Delta region,
and a number of new single-$Q^2$~\cite{Frolov,Mainz01,Bonn97} and
$Q^2$-dependent~\cite{Kamalov,Lee} fits have been carried out, in
the hope that a better determination of $\Delta (1232)$ properties
might now be possible.

These recent determinations have generally confirmed that the E2/M1
ratio remains ``small" (compared to the PQCD limit of 100\%) at
moderate values of $Q^2$.  However, while some fits~\cite{Kamalov}
find a cross-over to positive values, below 5~\GeVc, others do
not~\cite{Frolov,Lee}.  For this reason, we have made a number of
fits, both $Q^2$-dependent and single$-Q^2$, in order to see if a
clear trend emerges.

In the following section, we will give an overview of the existing
data, and indicate where the abovementioned Jefferson Lab
measurements will be added.  We will then briefly outline the
methods used in our fits.  Results for the E2/M1 and S2/M1 ratios
will be compared to other recent determinations.  Finally, we will
attempt to draw some conclusions from this exercise.

\begin{figure}[t!]
\centerline{%
\epsfig{file=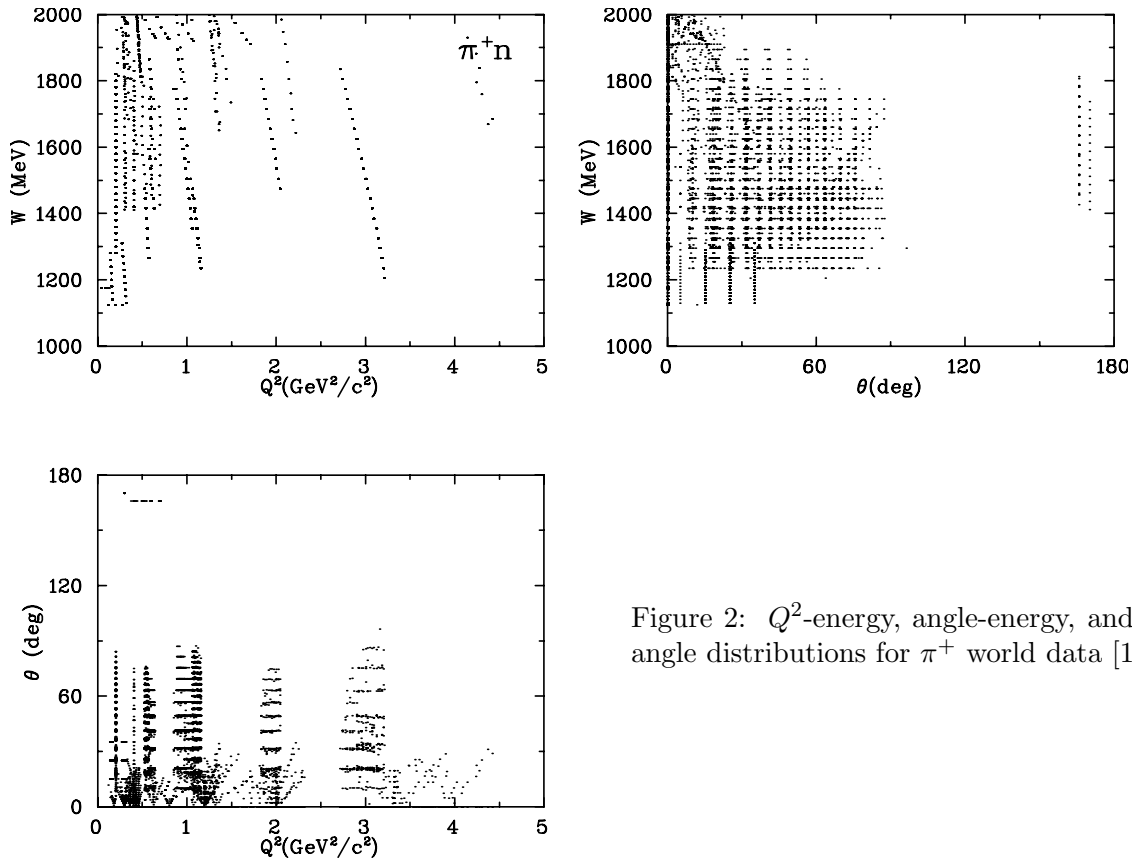,height=.9\textwidth,clip=,silent=,angle=90}}
\vspace{-5mm}
\hspace*{.55\textwidth}\raisebox{30mm}[0pt][0pt]{\parbox{.45\textwidth}{\caption[fig2]{\label{fig2}
$Q^2$-energy, angle-energy, and $Q^2$-angle distributions
for $\pi^+$ world data \protect\cite{SAID}.}}}
\end{figure}

\section{The Database}

It is somewhat more involved to show the data-distribution in
electroproduction (a function of $W$, $Q^2$, $\theta$, $\phi$,
$\epsilon$) than photoproduction (a function of $W$, $\theta$).
In Figs.~\ref{fig1} and ~\ref{fig2}, we have given 3 projections
(with a sum over all $\epsilon$ and $\phi$ values) for both the
$\pi^0$ and $\pi^+$ datasets.  This serves to show much of the
database is limited to $Q^2$ values below about 1 \GeVc,
and how little is measured for $\pi^+$ electroproduction at
backward angles ($\theta$).  The preliminary CLAS data are
similarly concentrated below 2~\GeVc, but have much better
angular coverage in the $\pi^+$ channel.

\section{Fitting Strategy}

The method we have used to fit electroproduction data is a direct
modification of our photoproduction formalism.  As
in the photoproduction case, correct threshold behavior and
Watson's theorem are built in.  Multipoles are parameterized
using the form
\begin{equation}
M = (\mbox{\rm Born} + \alpha_B)(1+iT_{\pi N}) + \alpha_R T_{\pi N}
  + (\mbox{\rm Im} T - T^2)(\alpha_r^H + i \alpha_i^H)~,
\label{1}\end{equation}
wherein $T_{\pi N}$ is the $\pi N$ elastic T-matrix~\cite{GW_pin}
for the $\pi N$ partial wave connected to a particular multipole,
the Born term contains pion and vector-meson exchanges, and
$\alpha_B$, $\alpha_R$, $\alpha_r^H$, and $\alpha_i^H$ are
phenomenological structure functions.  At $Q^2 = 0$, this is the
form used in photoproduction.  Thus, our present photoproduction
analysis is used to anchor the fit at this point.  At non-zero
$Q^2$, the Born terms have built-in $Q^2$ dependence.  Other
terms were initially modified by a factor
\begin{equation}
f(Q^2) = { k\over { k(Q^2=0)}} {1\over {(1+ Q^2/0.71)^2}} 
         e^{-\Lambda Q^2} (1 + \alpha Q^2)~,
\label{2}\end{equation}
where $k$ is the photon CM momentum,
$\Lambda$ is a universal cutoff factor, and $\alpha$ is
searched for each multipole.  The fit was significantly
improved if further variability was allowed in the energy
dependence.  As a result, an additional parameter was
searched (constrained to zero at the resonant point $W_R$)
\begin{equation}
\alpha Q^2 \rightarrow Q^2 \left( \alpha + \beta \left[
{W\over W_R} - 1 \right] \right)~,
\label{3}\end{equation}
$W$ being the CM energy.

As in our photoproduction analysis, we have performed
energy/$Q^2$ dependent fits over the full kinematic range.
We have also fitted data clustered around particular $Q^2$
values. This allows us to look for trends or problems in
the global fit.

\begin{figure}[t!]
\epsfig{file=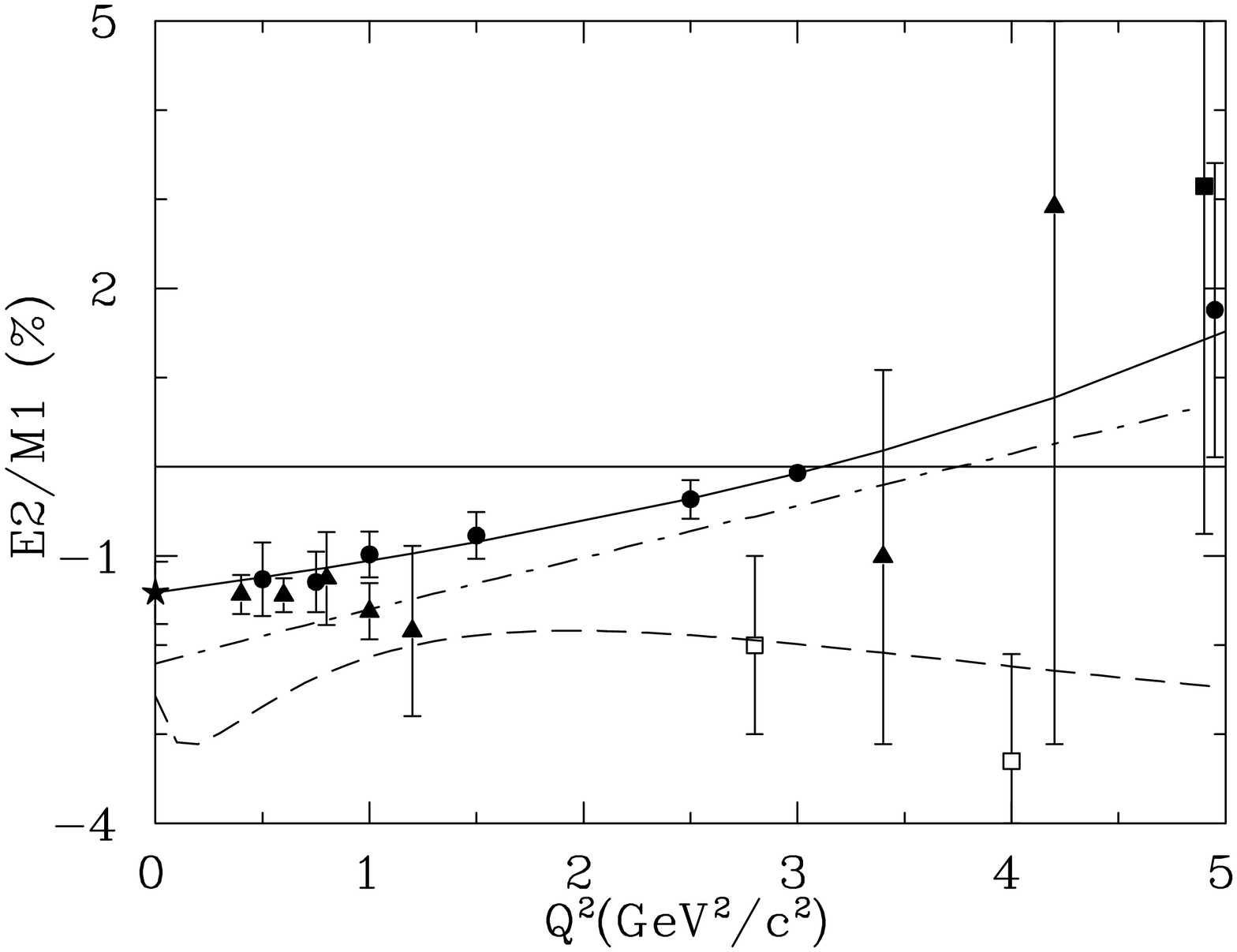,width=60mm,clip=,silent=,angle=90}
\hfill
\epsfig{file=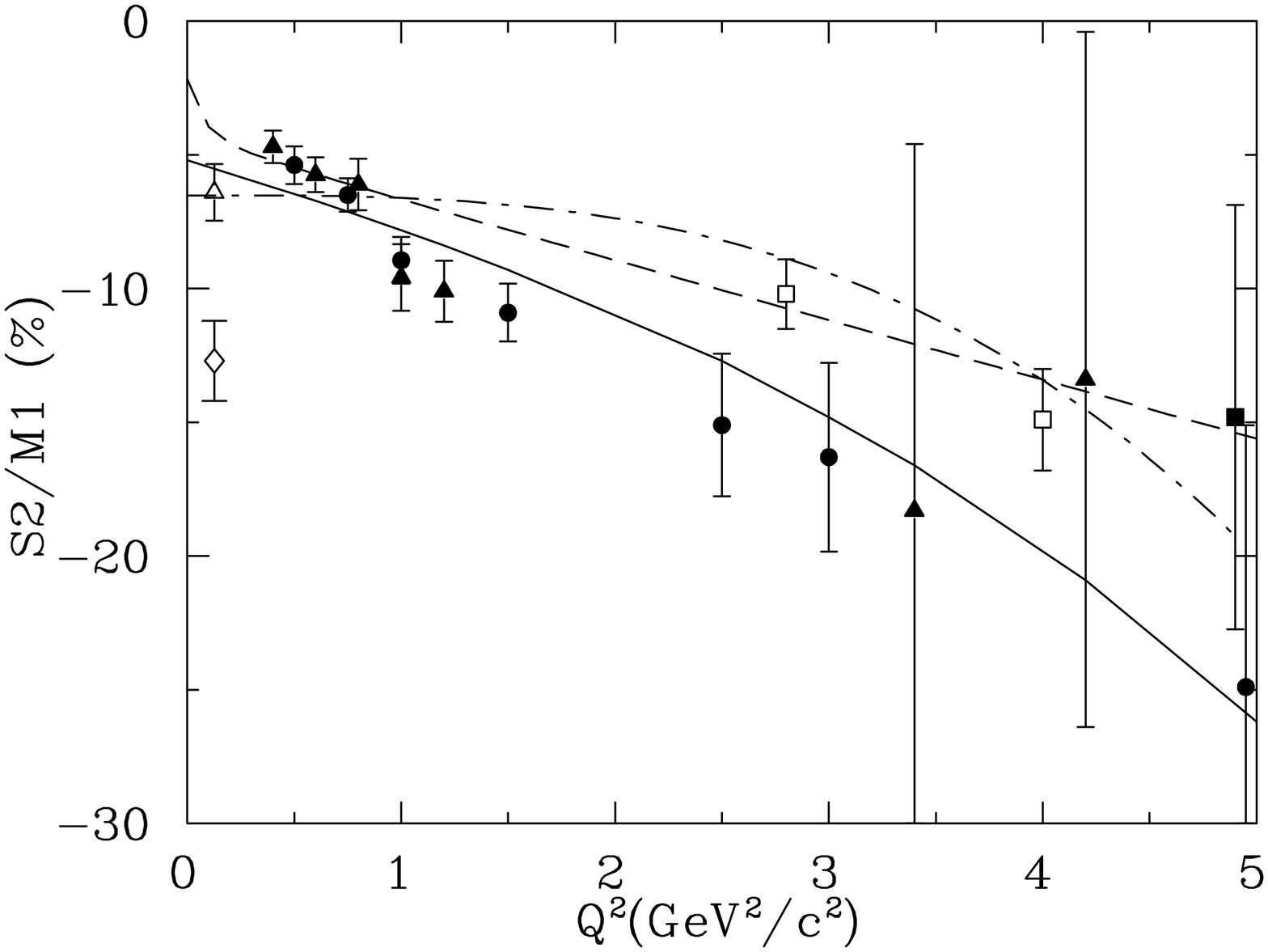,width=60.5mm,clip=,silent=,angle=90}
\hspace*{15mm} (a) \hspace{79mm} (b)\\[-4mm]
\caption[fig3]{\label{fig3}
   (a) E2/M1 and (b) S2/M1 ratios vs $Q^2$.  Values were extracted from our QDF (filled
   circles) using world and preliminary CLAS data
   (filled square: world data only) and SQS (filled triangle) solutions. Results from
   Ref. \protect\cite{Frolov}
   (open squares) are given in both (a) and (b). In addition, in (a),
   our pion photoproduction result
   ($Q^2$ = 0) \protect\cite{GW_photo} (filled asterisk),
   and in (b), the data of Refs. \protect\cite{Mainz01} (open triangle)
   and \protect\cite{Bonn97} (open diamond) are shown.
   The solid curves give best-fit results vs the set of QDF
   solutions.  Dash-dotted and dashed curves are from Refs.
   \protect\cite{Kamalov} and \protect\cite{Lee}, respectively.}
\end{figure}

\begin{table}[t!]
\begin{tabular}[t]{ccc}
QDF solutions:\\[5pt]
\hline
$Q^2_{\rm min} - Q^2_{\rm max}$ & \raisebox{-2mm}[0pt][0pt]{$\chi^2$/data} & \raisebox{-2mm}[0pt][0pt]{Data} \\
\GeVc  \\\hline
0.0$-$5.0  & 18713/10713 & NoCLAS \\ \hline
0.0$-$0.5  & 14647/9304  & CLAS   \\ \hline
0.0$-$0.8  & 32483/20734 & CLAS   \\ \hline
0.0$-$1.0  & 37610/24139 & CLAS   \\ \hline
0.0$-$1.5  & 48294/29820 & CLAS   \\ \hline
0.0$-$2.5  & 50013/31091 & CLAS   \\ \hline
0.0$-$3.0  & 51572/31837 & CLAS   \\ \hline
0.0$-$5.0  & 53828/33209 & CLAS   \\ \hline
\end{tabular}
\hfill
\begin{tabular}[t]{ccc}
SQS solutions:\\[5pt]
\hline
$Q^2_{\rm min} - Q^2_{\rm max}$ & \raisebox{-2mm}[0pt][0pt]{$\chi^2$/data} & \raisebox{-2mm}[0pt][0pt]{Data}  \\
\GeVc  \\\hline
0.2$-$0.4  & 10808/6733  & CLAS \\ \hline
0.4$-$0.6  & 17163/11020 & CLAS \\ \hline
0.6$-$0.8  &  9882/7497  & CLAS \\ \hline
0.8$-$1.0  &  4393/3274  & CLAS \\ \hline
1.0$-$1.2  &  8393/4529  & CLAS \\ \hline
2.8$-$3.4  &  1318/948   & CLAS \\ \hline
3.8$-$4.2  &  831/697    & CLAS \\ \hline
\end{tabular}
\caption{\label{tbl}
Comparison of $Q^2$-dependent (QDF) and single-$Q^2$
(SQS) solutions.}
\end{table}

\section{Comparisons and Conclusions}

Our results for $Q^2$-dependent (QDF) and single-$Q^2$ (SQS)
fits are summarized in Table~\ref{tbl}, for cases including
(CLAS) and excluding (NoCLAS) a set of preliminary CLAS data.
Results for the E2/M1 and S2/M1 ratios, as functions of $Q^2$,
are also displayed in Fig.~\ref{fig3}.  As is
evident from Table~\ref{tbl}, and Figs.~\ref{fig1} and
\ref{fig2}, the database is very sparse above a $Q^2$ of
about 1~\GeVc.  The single-$Q^2$ points in this region
have correspondingly large uncertainties.

Most fits were carried out including preliminary CLAS
data.  A single fit over the full range, excluding the
CLAS data, is given for comparison.  In Fig.~\ref{fig3},
all variants of the fit are included,
along with the fits of Refs.~\cite{Frolov,Mainz01,Bonn97}
and the analyses of Refs.~\cite{Kamalov,Lee}.  Our global
fit tends to follow more closely the result of
Ref.~\cite{Kamalov}.  Our single-$Q^2$ fits, though
confirming the trend seen in the global fits, have such
large uncertainties that they actually overlap with the
ratios extracted in Ref.~\cite{Frolov}.  An improvement
will require a more complete coverage of measurements
above 1$-$2~\GeVc.


\acknowledgments{This work was supported in part by a U.S.~Department of
Energy Grant No.~DE-FG02-99ER41110.  We also acknowledge
a contract from Jefferson Lab under which part of this
work was done. Jefferson Lab is operated by the
Southeastern Universities Research Association under DOE
contract DE-AC05-84ER40150.
}


\end{document}